\documentstyle[preprint,aps,epsf]{revtex}

\begin{document}

\title{Nuclear processes in magnetic fusion reactors with polarized fuel}
\author{ Michail P. Rekalo \footnote{ Permanent address:
\it National Science Center KFTI, 310108 Kharkov, Ukraine}}
\address{Middle East Technical University, Physics Department, Ankara 06531, 
Turkey\\
} 
\author{Egle Tomasi-Gustafsson}
\address{DAPNIA/SPhN, C.E.A./Saclay,  91191 Gif-sur-Yvette Cedex, France} 


\maketitle
\begin{abstract}
We consider the processes $d +d \rightarrow n +{^3\!He}$, 
$d +{^3\!He} \rightarrow p +{^4\!He}$, $d +{^3\!H} \rightarrow n +{^4\!He}$, 
${^3\!He} +{^3\!He}\rightarrow p+p +{^4\!He}$,   
${^3\!H} +{^3\!He}\rightarrow d +{^4\!He}$, with particular attention for 
applications in fusion reactors. After a model independent parametrization of 
the spin structure of the matrix elements for these processes at thermal 
colliding 
energies, in terms of partial amplitudes, we study 
polarization phenomena in the 
framework of a formalism of helicity amplitudes. 
The strong angular dependence of the final nuclei and of the polarization 
observables on the polarizations of the fuel components can be helpful in 
the design of the reactor shielding, blanket arrangement etc..We analyze also 
the angular 
dependence of the neutron polarization for 
the processes $\vec d +\vec d \rightarrow n +{^3\!He}$ and 
$\vec d +\vec {^3\!H} \rightarrow n +{^4\!He}$.
\end{abstract}


\narrowtext
\section{Introduction}
Nuclear fusion reactions, like  $d +d \rightarrow n 
+{^3\!He}$, or $d +{^3\!H} \rightarrow n +{^4\!He}$, are characterized by a 
large dependence on the spins of the colliding particles.  
It has been suggested \cite{Ku82} to use this property in magnetic fusion 
reactors with polarized nuclear fuel. A magnetic field of about 1 kG can keep 
the necessary direction of the polarization of the interacting nuclei, during a 
time which is longer in comparison with the reaction time.
Different technical solutions might be used: injection of polarized frozen 
pellets, or polarized targets for inertial fusion.

The strong dependence of the fusion reaction rates on the polarization states
results in an increasing or a decreasing of the cross section (with respect to 
the unpolarized case), depending on the colliding nuclei polarization 
directions. These characteristics can be used to optimize a fusion reactor in 
different ways:
\begin{itemize}
\item
The possible enhancement of the fusion rates for $\vec d+\vec{^3\!He}$ and the 
suppression of $\vec d+\vec d$-collisions would make this fuel competitive with 
$d+{^3\!H}$, as it would, in particular,  result in a {\it clean} reactor.
\item The strong anistropy of the neutron angular dependence in $\vec 
d+\vec{^3\!H}$-collision helps in optimizing the reactor shielding and the 
blanket 
design.
\item $\vec d+\vec {^3\!H}$-collisions can be source of intensive monochromatic 
polarized neutrons, with the choice of the polarization direction.
\end{itemize}

A precise knowledge of the spin structure of the threshold matrix elements for 
the processes induced by: $ d+ {^3\!He}$, $d+{^3\!H}$, ${^3\!He}+ {^3\!He}$, 
${^3\!H}+ ^3\!He$-collisions is required. At energies up to 10 keV, which are 
typical for fusion 
reactors, the S-state interaction of the colliding particles dominates and the   
general analysis of polarization phenomena is essentially simplified.

For this particular low energy domain, the general analysis of polarization 
phenomena \cite{Ke74} (in terms of scalar or helicity amplitudes) does not 
seem adequate, as it does not include the simplified characteristics of 
threshold regime. The main reason is that, at threshold, there is only one 
physical direction: the 3-momentum of the produced particles. Therefore, the 
parametrization of the corresponding matrix elements has to be done on the basis 
of one direction. It can not be derived as a limiting case of a general approach
based on two independent directions, the 3-momenta of the initial and final 
particles, which define the reaction plane. For S-state collisions we can not 
define this plane, having only one direction. The analysis of threshold 
polarization phenomena is also simpler and requires a dedicated parametrization, 
which is not a limiting case of a general parametrization. 

Note that, in this connection, the spin structure for the threshold kinematics 
is equivalent to (and sometimes simpler than) the case of collinear kinematics.

Our aim is to analyze here in the most general and complete form the reactions 
relevant to magnetic fusion reactors, with polarized fuel. Following 
the line of our previous paper \cite{Rek98}, where we considered  the reaction
$d +{^3\!He} \rightarrow p +{^4\!He}$, we will give here the general 
parametrization of the threshold amplitudes for $d +{^3\!He}$, $ d+d$ and 
${^3\!He}+{^3\!He}$-collisions, with special 
attention to the angular distribution of the reaction products for different 
possible polarization states of the colliding particles, without any particular 
assumption about the reaction mechanism.  For this aim we develop a formalism 
for the parametrization of the spin structure of  the threshold matrix elements 
in terms of the small number of allowed partial amplitudes, taking into account 
the most general 
properties of the strong interaction. 

In a fusion reactor the reaction rates and the 
angular distributions depend on  
the direction of the magnetic field. We use in this analysis a particular set of 
helicity amplitudes, with quantization axis along the direction of the magnetic 
field. (The effects of the magnetic field were not discussed in our previous 
paper \cite{Rek98}). 

The paper is organized as follows. In section 2 on the basis of some of our 
previous 
results concerning  the reaction $d +{^3\!H} ({^3\!He}) \rightarrow n (p) 
+{^4\!He}$, we derive the angular dependence of the differential cross 
sections for different polarization states of the colliding particles, and the 
angular dependence of the polarization of the produced neutrons (protons).

The process  $d +d \rightarrow n +{^3\!He}$, discussed in Section 3, is 
characterized by a set of three independent partial amplitudes (for S-state 
$dd$-collisions). We derive the limits of the 
integral coefficients for polarized particles collisions, which quantify the 
change of the cross section due to the 
polarizations of the  colliding particles. These coefficients depend only on the 
ratio of the square of the 
partial amplitudes.

Section 4 contains the discussion of the properties of some processes induced by
${^3\!He} +{^3\!He}$, ${^3\!H} +{^3\!H}$, and ${^3\!H} +{^3\!He}$-collisions.
The Pauli principle (for colliding or produced identical fermions), 
essentially simplifies the spin structure of these reactions, and it is 
possible in some cases to give model-independent predictions for 
polarization phenomena.
\section{\boldmath{The complete experiment for the reaction $\lowercase{d} 
+^3\!H (^3\!H\lowercase{e}) \rightarrow \lowercase{n (p)} +^4\!H\lowercase{e}$} 
}
\subsection{Introductory remarks}

The reaction $d +^3\!H \rightarrow n +^4\!He$ in the near threshold region is 
very interesting for the production of thermonuclear energy and plays an 
important role in primordial nucleosynthesis.The  lowest 
$\displaystyle\frac{3}{2}^+$ level of ~$^5\!He$ 
has excitation energy $E_x=16.75$ MeV (only 50 keV above $d +^3\!H$-threshold) 
and has a width of 76 keV. 

The microscopic explanation of the nature and the properties of this resonance 
is very complicated and still under debate in the physics of light nuclei.
The interpretation \cite{Br87} of this resonance as a shadow pole \cite{Ed64} 
introduces a new concept in nuclear physics, after atomic and particle physics. 
The possibility that the corresponding shadow poles for the 
two charge symmetric systems $d +^3\!He$ and $d +{^3\!H}$ (or 
$p +{^4\!He}$ and $n +{^4\!He}$) occupy different Riemann sheets, due to the 
difference in electric charges of the participating particles, can not be 
presently ruled out. Such phenomena can be considered as a new mechanism of 
violation of isotopic invariance of the strong interaction \cite{Cs93}.

Due to the close connection of the three processes $ d +{^3\!He} \rightarrow d 
+{^3\!He},~n +{^4\!He} \rightarrow n +{^4\!He}$ and $d +{^3\!H} \rightarrow n 
+{^4\!He}$ through the unitarity condition, the partial wave analysis 
\cite{Ho66,Je80} can not be performed independently for each reaction. The 
corresponding amplitudes are complex functions of the excitation energy. 
The multilevel ${\cal R}-$matrix approach allows to parametrize this dependence 
in terms of few parameters as shift, penetration factors and hard-sphere phase 
shift 
\cite{La58}. All characteristics of the ${\cal 
J}^P=\displaystyle\frac{3}{2}^+$-resonance, like the position, the width and 
particularly the Rieman sheet, can be found using an $S-$matrix approach
\cite{Br87,Pe89,Cs97,Ba97}.

The polarization phenomena are very important in the near threshold region, even 
for the S-state interaction. In this respect the reaction $d +{^3\!H} 
\rightarrow n +{^4\!He}$ plays a 
special role, because the presence of a D-wave in the final state results in 
nonzero one-spin polarization observables, such as, for example, the tensor 
analyzing power. In order to fully 
determine the two possible threshold (complex) amplitudes, two-spin polarization 
observables have to be measured, for example in collisions of polarized deuteron 
with polarized ${^3\!He}$-target. Here we will generalize our previous analysis 
\cite{Rek98}, taking into account the presence of  a magnetic field, which is 
necessary in order to conserve the 
polarization of the fuel constituents in a magnetic fusion reactor \cite{Ku82}.

For very small colliding energies the 
analysis of polarization 
phenomena for the reaction $d +{^3\!H} \rightarrow n +{^4\!He}$ can be carried 
out in a general form.
In the framework of a formalism, based on the polarized structure functions, we 
will point out the observables which have to be measured in order to 
have a full reconstruction of the spin structure of the threshold amplitudes. 
Data on cross section and tensor 
analyzing power exist, at threshold \cite{Dr80} (for a review see \cite{Ti98}).  
Among the two-spin observables, the measurement of a spin correlation 
coefficient, together with the cross section and the tensor analyzing power, 
allows to realize the complete experiment.

\subsection{Spin structure of the matrix element}
Let us first establish 
the spin structure of the matrix element. From the P-invariance of the strong 
interaction and the conservation of the total angular momentum, two partial 
transitions, for  $d +{^3\!He} \rightarrow p +{^4\!He}$ (as well as for $d 
+{^3\!H} \rightarrow n +{^4\!He}$) are allowed:
\begin{eqnarray}
&S_i=\displaystyle\frac{1}{2}~\rightarrow {\mathcal 
J}^{\pi}=\displaystyle\frac{1}{2}^+~\rightarrow {\ell}_f=0,\nonumber\\
&S_i=\displaystyle\frac{3}{2}~\rightarrow {\mathcal  
J}^{\pi}=\displaystyle\frac{3}{2}^+~\rightarrow {\ell}_f=2,
\end{eqnarray}
where $S_i$ is the total spin of the $ d+^3\!He$-system and ${\ell}_f$ is the 
orbital angular momentum of the final proton. The spin structure of the 
threshold matrix element can be parametrized in the form:
\begin{equation}
\begin{array}{ll}
&{\mathcal M}={\chi}_2^{\dagger}{\mathcal F}_{th} {\chi}_1,\nonumber\\
&{\mathcal F}_{th} =g_s\vec \sigma\cdot \vec D+ g_d (3\vec k\cdot\vec D~
\vec \sigma\cdot \vec k-\vec \sigma\cdot \vec D),
\end{array}
\end{equation}
where $\chi_1$ and $\chi_2$ are the two component spinors of the initial 
$^3\!He$ and final $p$, $\vec D$ is the 3-vector of the deuteron polarization 
(more exactly, $\vec D$ is the axial vector due to the positive parity of the 
deuteron), $\vec k$ is the unit vector along the 3-momenta of the proton (in 
the CMS of the considered reaction) and $g_s$ and $g_d$ are the amplitudes of 
the $S-$ and $D-$ production of the final particles, and they are complex 
functions of the excitation energy. Note that, in the general case, the spin 
structure of the matrix element, for the considered processes, contains six 
different contributions and the corresponding amplitudes are 
functions of two variables. 

The general parametrization of the differential cross 
section in terms of the 
polarizations of the colliding particles (in S-state) is given by:
\begin{eqnarray}
&\displaystyle\frac{d\sigma}{d\Omega}(\vec d+\vec {^3\!He} )=
\left (\frac {d\sigma}{d\Omega}\right )_0\left[\right .&1+ {\mathcal 
A}_1(Q_{ab}k_ak_b)+
{\mathcal A}_2\vec S\cdot\vec P\nonumber\\
&&+\left. {\mathcal A}_3\vec k\cdot\vec P~\vec k\cdot\vec S+
{\mathcal A}_4\vec k\cdot\vec P\times\vec Q\right ],
~Q_a=Q_{ab}k_b,
\end{eqnarray}
where $(d\sigma/d\Omega)_0$ is the differential cross section with unpolarized 
particles, $\vec P$ is the axial vector of the target ($^3\!He$) polarization, 
$\vec S$ and 
$Q_{ab}$ are the vector and tensor deuteron polarizations. The density matrix of 
the deuteron can be written as:
\begin{equation}
\overline{D_aD_b^*}=\displaystyle\frac{1}{3}(\delta_{ab}-\frac{3}{2}i\epsilon_{a
bc}S_c-Q_{ab}),~~Q_{aa}=0,~~Q_{ab}=Q_{ba}.
\end{equation}

After summing over the final proton polarizations one can find the following 
expressions:
\begin{equation}
\begin{array}{rlrl}
{\mathcal A}_1\displaystyle\left (\frac {d\sigma}{d\Omega}\right 
)_0=&-2{\mathcal 
R}e~g_sg_d^*-|g_d|^2,&
{\mathcal A}_2\displaystyle\left (\frac {d\sigma}{d\Omega}\right 
)_0=&-|g_s|^2-{\mathcal 
R}e~g_sg_d^*+2|g_d|^2,\nonumber\\
{\mathcal A}_3\displaystyle\left (\frac {d\sigma}{d\Omega}\right )_0=&3{\mathcal 
R}e~g_sg_d^*-3|g_d|^2,&
{\mathcal A}_4\displaystyle\left (\frac {d\sigma}{d\Omega}\right 
)_0=&-2{\mathcal 
I}m~g_sg_d^*.
\end{array}
\end{equation}
The coefficients ${\mathcal A}_i$ are related by the following linear 
relation: ${\mathcal A}_1+{\mathcal A}_2+{\mathcal A}_3=-1$ for any choice of 
amplitudes $g_s$ and $g_d$.
The integration of the differential cross section over the $\vec k$-directions 
gives:
$$\sigma(\vec d+\vec {^3\!H} )=\sigma_0(1+{\mathcal A}\vec S\cdot\vec 
P),~~{\mathcal 
A}={\mathcal 
A}_2+\displaystyle
\frac{1}{3}{\mathcal 
A}_3=\displaystyle\frac{-|g_s|^2+|g_d|^2}{|g_s|^2+2|g_d|^2},$$
and it is independent from the tensor deuteron polarization.

The presence of S-wave contribution (the amplitude $g_s$), decreases the value 
of the integral coefficient ${\mathcal A}$ whereas, in the fusion resonance 
region, 
where the D-wave dominates, the maximum value, 
${\mathcal A}=1/2$, is reached. In the 
complete experiment (which gives $|g_s|^2$, $|g_d|^2$ and 
$Re~g_sg_d^*$), the amplitudes $|g_s|$ and $|g_d|$ can be found in a model 
independent way, with the help of the following formulas: 
\[
9{|g_s|^2}=\left (5+2{\mathcal A}_1-4{\mathcal A}_2\right )\left 
(\displaystyle\frac 
{d\sigma}{d\Omega}\right )_0,
\]
\begin{equation}
9{|g_d|^2}=\left (2-{\mathcal A}_1+2{\mathcal A}_2\right )\left 
(\displaystyle\frac 
{d\sigma}{d\Omega}\right )_0,
\end{equation}
\[
-9{\mathcal R}e{g_s}{g_d^*}=\left (1+4{\mathcal A}_1+{\mathcal A}_2\right 
)\displaystyle\left (\frac {d\sigma}{d\Omega}\right )_0.
\]
One can see that the {\it 
integral} 
coefficient ${\mathcal A}$ can be determined from  polarized 
nuclei collisions by measuring:
\begin{itemize}
\item the tensor analyzing power ${\mathcal A}_1$ in  $\vec d +{^3\!He} 
\rightarrow p +{^4\!He}$,
\item the spin correlation coefficient $C_{xx}= C_{yy}={\mathcal A}_2$ (if the 
$z-$axis is 
along $\vec k$-direction.)
\end{itemize}

Let us study now the polarization properties of the outgoing nucleons.
We will show that can be predicted only from the tensor analyzing power, 
${\mathcal A}_1$. The polarization 
$\vec P_f$ of the produced nucleon depends on the polarization $\vec P$ of the 
initial ${^3\!He}$ (or ${^3\!H}$) as follows:
$\vec P_f=p_1\vec P+p_2\vec k~\vec k\cdot\vec P$, where the real coefficients 
$p_i,~i=1,2$, characterize the spin transfer coefficients (from the initial 
${^3\!He}$ or ${^3\!H}$ to the final nucleon): $K_x^{x'}=p_1+p_2\cos^2\theta 
,~K_x^{z'}=p_2~\sin\theta ~\cos\theta, $ where 
$\theta$ is the angle between $\vec k$ and $\vec P$.
Averaging over the polarizations of the initial deuteron, we can find:
$$
p_1\left (\frac {d\sigma}{d\Omega}\right )_0=-\displaystyle\frac{1}{3}\left (
|g_s|^2+4 Re~ g_sg_d^*+4|g_d|^2\right ),$$
$$
p_2\displaystyle\frac{d\sigma}{d\Omega}_0=4 Re~ g_sg_d^*+2|g_d|^2,
$$
$$3p_1=-1+2{\mathcal A}_1,~p_2=-2{\mathcal A}_1,~~3p_1+p_2=-1.$$
This analysis  holds in the presence of S-state only, in the entrance 
channel. The validity of this assumption can be experimentally verified with the 
measurement of T-odd one-spin
polarization observables, as the analyzing powers in $\vec d +{^3\!He} 
\rightarrow p +{^4\!He}$ induced by vector deuteron polarization or $d +\vec 
{^3\!He} \rightarrow p +{^4\!He}$. This observable is very sensitive to the 
presence of even a small P-wave contribution, due to its interference with the 
main amplitude. 

\subsection{Helicity amplitudes}
We calculate here the helicity amplitudes $ 
F_{\lambda_1\lambda_2,\lambda_3}$, with ${\lambda_1}={\lambda_d}$, 
${\lambda_2}={\lambda_{^3He}}$, ${\lambda_3}={\lambda_p}$ (or ${\lambda_n}$), in 
terms of the partial amplitudes $g_s$ and $g_d$. This formalism is very well 
adapted for the analysis of angular distributions of the reaction products, in 
conditions of fusion reactors (with polarized fuel) and to the description of 
polarization phenomena. The direction of magnetic field $\vec B$ can be chosen 
as 
the most preferable quantization axis ($z-$axis). The formalism of the helicity 
amplitudes allows to study the angular dependence of the polarization 
observables, relative to $\vec B$. For example, the polarization properties of 
the neutron in 
$\vec d +\vec {^3\!H} \rightarrow n +{^4\!He}$ can be easily calculated in terms 
of these 
amplitudes.

The  peculiar strong angular dependence of all observables 
is due to the presence (in conditions of fusion polarized reactor) of two 
independent physical directions, $\vec k$ and $\vec B$. So even for the S-state 
interaction, a non trivial angular dependence of the reaction products appears, 
i.e. some angular anisotropy, related to the initial polarizations. As all the 
polarizations of both colliding particles depend on the same magnetic field 
$\vec B$, the results for these observables depend only on the angle
$\theta$, between $\vec k$ and $\vec B$. The case of the collision of polarized 
beam with polarized target, where the beam and the target may have different 
directions of polarization is more complicated, but it can also be treated in 
the framework of the helicity formalism.

The deuteron polarization vector $\vec D^{(\lambda)}$ 
(with a definite helicity 
$\lambda$), can be chosen as :  $\vec D^{(0)}=(0,0,1)$ and $\vec 
D^{(\pm)}={1}/{\sqrt{2}}(\pm 1,i,0)$. So the following expressions for the 
six possible helicity amplitudes can be found:
\begin{equation}
\begin{array}{ll}
F_{0+,+}=g_s-(1-3\cos^2\theta)g_d,&F_{++,-}=\displaystyle\frac{3}{
\sqrt{2}}\sin^2\theta g_d,\nonumber\\
F_{0+,-}=\displaystyle\frac{3}{2}\sin~2\theta g_d,&
F_{-+,+}=\displaystyle\frac{3}{2\sqrt{2}}\sin~2\theta g_d,\nonumber\\
F_{++,+}=-\displaystyle\frac{3}{2\sqrt{2}}\sin~2\theta g_d,&
F_{-+,-}=-\frac{1}{\sqrt{2}}\left [ 2 g_s+ (1-3\cos^2\theta)g_d\right ],
\end{array}
\end{equation}
where $\theta$ is the nucleon production angle, relative to  the $\vec B$ 
direction. Other possible helicity amplitudes, with reversed helicities of all 
particles, can be obtained from (7), by parity reversion.

One can see, that for collinear kinematics, i.e. for $\theta=0^{\circ}$, only 
two helicity 
amplitudes are nonzero: $F_{0+,+}=g_s+2g_d,~~~F_{-+,-}=-\sqrt{2}(g_s-g_d).$ So 
for the corresponding differential cross section (for collisions of 
unpolarized particles) one can find:
$$
\left(\displaystyle\frac{d\sigma}{d\Omega}\right )_{\theta=0}=\frac{1}{3}\left 
(|F_{0+,+}|^2+|F_{-+,-}|^2\right )=|g_s|^2+2|g_d|^2.$$
The same formula is also correct in the general case (i.e., for $\theta\ne 0$)
for collisions of unpolarized particles:
$$
\left(\displaystyle\frac{d\sigma}{d\Omega}\right )_0=\frac{1}{6} \sum_{\lambda} 
|{\mathcal 
F}_{\lambda_1\lambda_2,\lambda_3}|^2= \left 
(\displaystyle\frac{d\sigma}{d\Omega}\right )_{\theta=0}.$$
This relation between the differential cross section for collinear kinematics 
and the cross section at any angle $\theta$ is valid for S-state interaction, 
only.
\subsection {Collision of polarized particles}
The angular dependence of the reaction 
products in $d +{^3\!H} \rightarrow n +{^4\!He}$ for different polarization 
states of the colliding particles can be  derived from (7).
\begin{itemize}
\item Collisions of longitudinally polarized deuterons ($\lambda_d=0$), with 
polarized $^3\!H$ or $^3\!He$: 
\end{itemize}
\begin{equation}
\sigma_{0+}(\theta)=|F_{0+,+}|^2+|F_{0+,-}|^2=|g_s|^2+ 
2Re~g_sg_d^*(-1+3\cos^2\theta)+|g_d|^2(1+3\cos^2\theta).
\end{equation}
\begin{itemize}
\item $\vec d+\vec{^3\!He}$ collisions with parallel polarizations (relative to 
$\vec B$):
\end{itemize}
\begin{equation}
\sigma_{++}(\theta)=|F_{++,+}|^2+|F_{++,-}|^2=\frac{9}{2}|g_d|^2\sin^2\theta.
\end{equation} 
\begin{itemize}
\item $\vec d+\vec{^3\!He}$ collisions with antiparallel polarizations:
\end{itemize}
\begin{equation}
\sigma_{+-}(\theta)=|F_{+-,+}|^2+|F_{+-,-}|^2=2|g_s|^2+2Re~g_sg_d^*
(1-3\cos^2\theta)+ \frac{1}{2}(1+3\cos^2\theta)|g_d|^2.
\end{equation}
The sum of all these polarized cross sections is independent from polar angle 
$\theta$: the unpolarized 
cross section is isotropic, as expected for $S-$state interaction.
 
For the pure fusion resonance (with $g_s=0$), the angular distribution of the 
reaction products depends specifically on the direction of the polarizations of 
the colliding 
particles: the $\sin^2\theta$-dependence for parallel (++) collisions, becomes a 
dependence in $(1+3\cos^2\theta)$ for (+-) and (0+) collisions, to be compared  
with the isotropic behavior of the  
unpolarized collisions. Such definite and strong anisotropy can play a very 
important role in the design of the neutron shield of a reactor and of the 
blanket, where energetic 
neutrons  
(from $d +{^3\!H} \rightarrow n +{^4\!He}$ ) can produce ${^3\!H}$ through the 
reaction $n +{^6\!Li}\rightarrow {^3\!H} +{^4\!He}$. Once a $d +{^3\!H} 
$-reactor is beginning to operate,   ${^3\!H}$-fuel can be produced  in 
$^6\!Li$-blanket. In principle, this blanket can contain polarized $\vec 
{^6\!Li}$, 
for a more efficient ${^3\!H}$-production in $\vec n+\vec {^6\!Li}$-collisions. 

From Figs. 1 and 2, one can see that the angular dependence of the cross 
sections for polarized collisions, is essentially influenced by the presence of 
the S-wave amplitude and its relative phase.

Let us calculate now the following ratios:
$$R_{\lambda_1\lambda_2}=\displaystyle\frac{\int_{-1}^{+1}\sigma_{\lambda_1
\lambda_2}(\theta )d\cos\theta}
{\int_{-1}^{+1}d\cos \theta (d\sigma/d\Omega)_0,}$$
from Eqs. (8-10) for $\sigma_{\lambda_1\lambda_2}(\theta)$:
\begin{equation}
R_{0+}=1,~~R_{++}=\frac{3}{2}f,~~~~R_{+-}=\displaystyle\frac{1}{2}(4-3f).
\end{equation}
So we can write the following limits:
$$0\le R_{++}\le 3/2~{(g_s=0)},~~1/2\le R_{+-}\le 2~(g_d=0).$$
In the fusion resonance region, ($f$=1)\footnote{In particular the ratio of 
amplitudes 
$f=2|g_d|^2/(|g_s|^2+2|g_d|^2)$ was 
firstly defined in 
\protect\cite{Ku82}.}, the 
(++)-collisions increase the reaction yield (in comparison with collisions of 
unpolarized particles) with a maximum coefficient $\le$ 3/2, for pure D-wave 
fusion 
resonance. 
Using the notations of \cite{Ku82} one can obtain the following 
general formula for the differential cross section of  $\vec d +\vec {^3\!H}$ 
(or $\vec d +\vec {^3\!He}$)-collisions:
$$\displaystyle\frac{d\sigma}{d\Omega}
(\vec d+\vec {^3\!H})=6|g_d|^2\left \{ \frac{3}{4}a \sin^2\theta+\frac{b}{6} 
\left [\frac{2}{f}-(1-3\cos^2\theta)\left (1+\displaystyle\frac{2Re~ 
g_sg_d^*}{|g_d|^2}\right)\right]+\right .
$$
\begin{equation}
\left .+\frac{c}{12}\left [ \frac{8}{f}
-6-(1-3\cos^2\theta )\left( 1-\displaystyle\frac
{4Re~g_sg_d^*}{|g_d|^2}\right )\right ]\right \}.
\end{equation}
Here $a=d_+t_++d_-t_-$, $b=d_0$, $c= d_+t_++d_-t_+$ and  $d_+$, $d_0$, $d_-$ are 
the fractions of deuterons with polarization respectively parallel, 
transverse, antiparallel to $\vec B$ , while  $t_+$ and $t_-$ are the 
corresponding fractions for ${^3\!H}$. The relations $d_++d_0+d_-$=1 and 
$t_++t_-=1$ hold. The 
case $a=b=c=1/3$ corresponds to unpolarized collisions.

Note that the predicted angular dependence for $b$ and $c$ contributions, Eq. 
(12), differs essentially from the corresponding expression of \cite{Ku82}. It 
coincides only for the special case $f=1$, $g_s=0$. The denominator in Eq. (2) 
from ref. \cite{Ku82} must  be also different.

From (12) one can find the following expression for the differential cross 
section of collisions of polarized deuterons with unpolarized ${^3\!H}$:
$$\displaystyle\frac{d\sigma}{d\Omega}(\vec 
d+{^3\!H})=2|g_d|^2\left[\frac{1}{f}+P_{zz}
\displaystyle\frac{1-3\cos^2\theta}{4}
\left(1+\displaystyle\frac{2Re~g_sg_d^*}{|g_d|^2}\right)\right],
$$
i.e. it depends on the tensor deuteron polarization only. We used above the 
standard definition: 
$P_{zz}=d_+-2d_0+d_-$. Due to the $ (1-3\cos^2\theta)$ dependence, 
after integration 
over $\theta$, the cross section, again, does not depend on the deuteron 
polarization. 
\subsection{\boldmath{Polarization of neutrons in 
$\vec {\lowercase{d}} + \vec {^3\!H}$ collisions }}

Using the helicity amplitudes (7) it is possible to predict also the angular 
dependence of the neutron polarization in $\vec d+\vec {^3\!H}\rightarrow n 
+{^4\!He}$, in the general case of polarized particle collisions:
$$
(n_+-n_-)\displaystyle\frac{d\sigma}{d\Omega}(\vec d+\vec{^3\!H})= 
\frac{9}{2}(d_-t_--d_+t_+)\sin^2\theta(1-2\cos^2\theta)|g_d|^2+$$
$$+d_0(t_+-t_-)\left[|g_s|^2-2(1-3\cos^2\theta)Re~g_sg_d^*+(1-15\cos^2\theta+18
\cos
^4\theta)|g_d|^2\right ]+$$
$$+ 
(d_+t_--d_-t_+)\displaystyle\frac{1}{2} \left 
[4|g_s|^2+4(1-3\cos^2\theta)Re~g_sg_d^*+(1-15\cos^2\theta+18\cos
^4\theta)|g_d|^2\right ],$$
where $n_{\pm}$ is the fraction of neutrons, polarized parallel and 
antiparallel to the direction of the magnetic field.

Let us write some limiting cases of this general formula:
\begin{itemize}
\item[(a)] Collisions of polarized deuterons with unpolarized ${^3\!H}$-nuclei:
$$
(n_+-n_-)\displaystyle\frac{d\sigma}{d\Omega}(\vec 
d+{^3\!H})=(d_+-t_-)\left[ |g_s|^2 
+(1-3\cos^2\theta)Re~g_sg_d^*-\right .
$$\begin{equation}
\left .
 (2-3\cos^2\theta)|g_d|^2\right ].
\end{equation}
\item[(b)] Collisions of unpolarized deuterons with polarized ${^3\!H}$-nuclei:
\end{itemize}
$$(n_+-n_-)\displaystyle\frac{d\sigma}{d\Omega}(d+\vec{^3\!H})=\displaystyle
\frac{t_--t_+}{3}\left
[|g_s|^2 +4(1-3\cos^2\theta)Re~g_sg_d^*\right .
$$\begin{equation}
\left . +2(2-3\cos^2\theta)|g_d|^2\right ].
\end{equation}
In the case of fusion resonance ($g_s=0$), these formulas reduce to:
$$
(n_+-n_-)\displaystyle\frac{d\sigma}{d\Omega}(\vec d+\vec 
{^3\!H})=\frac{9}{4}\sin^2\theta
(1-2\cos^2\theta)( d_+t_--d_+t_+)+ 
$$
\begin{equation}
\frac{1}{2}\left [d_0(t_+-t_-)+\frac{1}{2}( d_+t_--d_-t_+) 
(1-15\cos^2\theta+18\cos^4\theta)\right ].
\end{equation}

Averaging over the polarizations of $d$ (or $^3\!H$) one can find
particular expressions:
$$
(n_+-n_-)\displaystyle\frac{d\sigma}{d\Omega}(d+\vec{ ^3\!H})=\frac{1}{2}
(t_--t_+)(2-3 \cos^2\theta)
$$
and
$$
(n_+-n_-)\displaystyle\frac{d\sigma}{d\Omega}(\vec d+^3\!H)=-\frac{1}{3}( 
d_--d_+)(2-3\cos^2\theta).
$$
The angular dependence of most of these polarization observables is sensitive to 
the relative value of the $g_s$ and 
$g_d$ amplitudes, due to the $g_sg_d^*$-interference contributions. Of course, 
in the region of the fusion resonance the $g_d$ amplitude is dominant. However 
the 
temperature conditions, typical for a fusion reactor, correspond to collision 
energies lower than the energy of the fusion resonance. Even a small $g_s/g_d$ 
ratio can change the angular behaviour of the polarization observables. In Figs. 
3-4 we show, in a 3-dimensional plot, the dependence of the neutron polarization 
on the ratio $x=|g_s|/g_d|$ and on the production angle $\theta$ for three 
values of the relative phase 
$\delta$, $\delta=0,~\pi/2,~\pi$, for $\vec d+^3\!H$ and $ 
d+\vec{^3\!H}$-collisions.

The exact determination of the parameters $x$ and $\delta$, is crucial for 
thermonuclear 
processes. This is a reason to perform a complete experiment for this reaction 
as discussed earlier \cite{Rek98}. The important point is 
that  even at very low energies, where the spin structure is simplified, a 
complete experiment must include the scattering of polarized beam on polarized 
target. The full reconstruction of the threshold matrix elements requires this 
type of experiment.
\section{\boldmath{Processes $\lowercase{d+d\rightarrow n+}^3\!H\lowercase{e}$ 
and 
$\lowercase{d+d\rightarrow p+}^3\!H$}}
\subsection{Introductory remarks}
The $d+d\rightarrow n+^3\!He$ and $d+d\rightarrow p+^3\!H$ reactions at low 
energy have a very wide spectrum of fundamental and practical applications, from 
the discovery of tritium and helium isotopes \cite{Ol33}, to the 
important role for primordial nucleosynthesis in the early Universe and fusion 
energy production with 
polarized and unpolarized fuel {\cite{Ku82,Ku86}. These processes are of large 
interest in nuclear theory: for example, in a four nucleon system, contrary to 
three nucleon system, broad resonant states can be excited \cite{Ti92}.
The angular dependence of the differential cross sections \cite{Br90,Kr87} and 
the polarization observables [11-14] for these charge symmetric reactions 
constitutes a good test of the isotopic invariance for the low energy nuclear 
interaction.
The $dd-$ interaction is also connected to muon catalyzed processes ($\mu d 
d)\rightarrow \mu+p+^3\!H$ or $(\mu dd)\rightarrow \mu +n+^3\!He$ \cite{Br89}, 
where only the P-state of the $dd$-system is present, at low energy. 

In the general case the spin structure of the matrix element for  
$d+d\to n+^3\!He~(p+^3\!H)$ is quite complicated, with 
18 independent spin combinations, and therefore with 18 complex scalar 
amplitudes, which are functions of the excitation energy and the scattering 
angle. However, at very small collision energies, where the 
$S-$state deuteron interaction has to dominate, this structure is largely 
simplified. The identity of the colliding deuterons, which are bosons, 
is an important guide for the partial amplitude analysis in order to determine  
the spin structure of the reaction amplitude. The determination of the 
polarization observables is indispensable, for this purpose. The four possible 
analyzing powers for $\vec d+d$-collisions, $A_y$, $A_{zz}$, $A_{xz}$ and 
$A_{xx}-A_{yy}$ were measured at $E_d\le100$ keV, as well as the angular 
dependence of the differential cross section \cite{Br90,Kr87,Fl94}.

The knowledge of the relative role of different orbital angular momenta (and 
 the corresponding partial amplitudes) is essential for the solution of 
different 
fundamental problems concerning these processes, like the possibility to build a 
thermonuclear "clean" reactor with polarized $d+^3\!He$-fuel. The main reaction 
$d+^3\!He\rightarrow p+^4\!He$ does not produce radioactive nuclei, and the 
possibility to decrease the cross section of $\vec d+\vec d$-collisions (which 
produces $n+^3\!He$ or $p+^3\!H$) with 
parallel polarizations, will decrease the production of neutrons and the 
tritium.
Direct experimental data about $\vec d+\vec d$- low energy collisions are 
absent, so the dependence of the cross section on the polarization states of the 
colliding particles can be  calculated  only from theoretical predictions or 
from 
different multipole analysis.

The theoretical predictions and the results of multipole analysis seem very 
controversial now, even at very low 
energy.
In the first partial wave analysis \cite{Ad69,Ad81} it was found that the 
S-state $dd-$interaction in the quintet state (i.e. with total spin $S_i=2$) is 
smaller in comparison with the $S_i=0$ interaction. This was consistent with the 
conclusion of ref. \cite{Ku82}, that in a polarized reactor it is possible to 
suppress $\vec d+\vec d$-collisions. Later \cite{Ho84}, it was pointed out that 
strong central forces with $D-$state in $^3\!He$ can induce a large $dd-$ 
interaction in the quintet state and resonating-group calculations \cite{Ho84} 
found that polarized collisions are not suppressed. On the other hand, DWBA 
calculations give a large suppression for the ratio of polarized on unpolarized 
cross section, $\sigma_{++}/\sigma_0\simeq 0.08$ in the range $E_d=20-150$ keV, 
even after inclusion of the $^3\!He$ D-state. A more recent analysis 
\cite{Le90,Pa92} based on $R-$matrix approach, concludes that this ratio does 
not decrease with energy. Note that in principle, it can be energy dependent 
\cite{Zh95}.

Again, a direct measurement of polarized $dd$-collisions would greatly help in 
solving 
these problems and the complete experiment will allow to reconstruct the spin 
structure of the reaction amplitude.  Therefore, the considerations based on 
$S-$wave only, have to 
be considered as the first necessary step which can illustrate the possible 
strategy of the complete experiment for this case.

\subsection{Partial amplitudes}
We establish here the spin structure of the threshold matrix element for the 
$d+d\rightarrow n+^3\!He ~(p+^3\!H)$ process. For S-state $dd-$interaction the 
following 
partial transitions are allowed:
\begin{eqnarray*}
&S_i=0~\rightarrow {\mathcal 
J}^{\pi}=0^+~\rightarrow S_f=0,~{\ell}_f=2,\\
&S_i=2~\rightarrow {\mathcal 
J}^{\pi}=2^+~\rightarrow S_f=0,~{\ell}_f=2,\\
&S_i=2~\rightarrow {\mathcal 
J}^{\pi}=2^+~\rightarrow S_f=1,~{\ell}_f=2,
\end{eqnarray*}
where $S_i$ is the total spin of the colliding deuterons, ${\ell}_f$ is the 
orbital angular momentum of the final nucleon. Note that the Bose statistics for 
identical deuterons allows only even values of initial spin, that is 
$S_i=0$ and $S_i=2$ for the S-state. The resulting  spin structure of the 
threshold matrix element can be written as:
\begin{equation}
\begin{array}{ll}
{\mathcal M}=i({\chi}_3^{\dagger}\sigma_2 \widetilde{{\chi}_1^{\dagger}})&\left[ 
g_1\vec 
D_1\cdot \vec D_2+g_2( 3 \vec k\cdot\vec D_1~\vec k\cdot \vec D_2-\vec D_1\cdot 
\vec D_2)\right .\nonumber\\
&\left .+g_3(\vec \sigma\cdot \vec k\times\vec D_1~\vec k\cdot \vec D_2+ 
\vec \sigma\cdot \vec k\times\vec D_2~\vec k\cdot \vec D_1) \right ],
\end{array}
\end{equation}
where $\chi_1$ and $\chi_3$ are the 2-component spinors of the produced 
nucleon and 
$^3\!He$ (or $^3\!H$), $\vec D_1$ and $\vec D_2$ are  the 3-vectors of the 
deuteron polarization, $\vec k$ is the unit vector along the 3-momenta of the 
nucleon (in 
the CMS of the considered reaction). The amplitudes $g_1$ and $g_2$ describe 
the production of the singlet $n+^3\!He$-state, and the amplitude $g_3$- the 
triplet state. The complete experiment in $S-$state $dd$-interaction implies the 
measurement of 5 different 
observables, to determine 3 moduli and two relative phases of partial 
amplitudes.

The validity of the S-state approximation in the near threshold region can be 
checked by measuring any T-odd polarization observable, the simplest of which 
are the one-spin observables as the vector analyzing power in the reaction $ 
\vec d+d \rightarrow n+^3\!He$ \cite{Fl94}. Note that Eq. (16) is correct also 
for the threshold matrix elements of the inverse process: 
$n+^3\!He\rightarrow d+d $ (or $p+^3\!H\rightarrow  d+d $). 
\subsection{Helicity amplitudes}
In order to establish the angular dependence of the reaction products, for 
collisions of polarized particles, in the presence of magnetic field, let us 
derive 
the  helicity amplitudes.  The spin structure of the $d+d$ reactions is more 
complex 
in 
comparison to $d+^3\!He$. The analysis of polarization phenomena is also more 
complicated. It was mentioned in \cite{Ku82}, that an enhancement factor, equal 
to 2 
can be obtained in a polarized plasma \footnote{Note that this holds only for 
the partial wave analysis \cite{Ad69,Ad81}.}, for the reaction $d+d\rightarrow 
n+^3\!He$, if 
the deuterons are polarized transversally to the direction of the magnetic 
field, i.e. 
for (00)-collisions, in an ordinary thermal ion distribution. Alternatively, if 
colliding beams or beam and target methods are used (inertial fusion), the two 
ions 
should 
be polarized in opposite direction, relatively to the field. In case of 
collisions of 
deuterons with parallel polarizations i.e (++) or ($--$), a large suppression of 
the 
reaction rate is expected.

It is then interesting to analyze all possible configurations of the 
polarization of 
the colliding deuterons. We can classify the helicity amplitudes according to 
the 
following scheme:
\begin{description}
\item [ I)]\underline{00~collisions}: the polarization is transverse to the 
magnetic 
field 
$\rightarrow$ 2 independent amplitudes;
\item [ II)]\underline{++~collisions}: the polarization parallel to the magnetic 
field 
$\rightarrow$ 4 independent amplitudes;
\item [III)]\underline{+$-$~collisions}: collisions with deuterons with 
antiparallel 
polarization, in the same direction as the magnetic field $\rightarrow $ 4 
independent 
amplitudes;
\item [IV)] \underline{0+~collisions}: collisions of one  deuteron with  
polarization 
transverse to the magnetic field with the other deuteron polarized along  the 
magnetic 
field $ \rightarrow $ 4 independent amplitudes;
\end{description}

The corresponding helicity amplitudes ${\mathcal 
F}_{\lambda_1\lambda_2,\lambda_3\lambda_4}$, (with ${\lambda_1}\equiv 
\lambda_{d_1}$, 
${\lambda_2}\equiv\lambda_{d_2}$, ${\lambda_3}\equiv \lambda_{^3\!He}$, 
${\lambda_4}\equiv \lambda_N$) are 
given in 
terms of partial amplitudes:
\vspace{.2truecm}

\noindent {\bf (I)}~~${\mathcal F}_{00,++}=-\sin~2\theta g_3, ~~~~{\mathcal 
F}_{00,+-}=g_1-(1-3\cos^2\theta)g_2,$

\noindent {\bf (II)}~~${\mathcal F}_{++,++}=\sin~2\theta g_3,~~~~{\mathcal 
F}_{++,+-}=\sin^2\theta
(\displaystyle\frac{3}{2}g_2+g_3),$

\noindent ~~~~~~$
{\mathcal F}_{++,--}=0,~~~~{\mathcal F}_{++,-+}=\sin^2\theta
(-\displaystyle\frac{3}{2}g_2+g_3),$
\begin{equation}
\mbox{\boldmath \bf (III)}~~{\mathcal F}_{+-,++}={\mathcal 
F}_{+-,--}=-\displaystyle\frac{1}{2}\sin~2\theta 
g_3,\hspace {5 true cm}
\end{equation}
\noindent ~~~~~~~~$
{\mathcal F}_{+-,+-}=-{\mathcal F}_{+-,-+}=-g_1-  
\displaystyle\frac{1}{2}(1-3\cos^2\theta)g_2.$

\noindent {\bf (IV)~~}$
{\mathcal F}_{0+,++}=\displaystyle\frac{1}{\sqrt{2}}(-1+3\cos^2\theta) 
g_3,~~~{\mathcal F}_{0+,+-}=\displaystyle\frac{1}{2\sqrt{2}}\sin~2\theta 
(3g_2+g_3),$

\noindent ~~~~~~~~$
{\mathcal F}_{0+,--}=-\displaystyle\frac{1}{\sqrt{2}}\sin^2\theta g_3, ~~
{\mathcal F}_{0+,-+}=\displaystyle\frac{1}{2\sqrt{2}}\sin~2\theta 
(-3g_2+g_3).$

\noindent where $\theta$ is the nucleon production angle relative to $\vec B$ 
direction.
\subsection{Angular dependence for collisions of polarized deuterons}
After summing over the polarization states of the produced particles,  the cross 
section of the process $\vec d+\vec d\rightarrow n+^3\!He$, for definite 
deuteron polarizations, can be written as:
$$\sigma_{00}(\theta)=2\left (|{\mathcal F}_{00,++}|^2+|{\mathcal 
F}_{00,+-}|^2\right 
)=2|g_1-g_2(1-3\cos^2\theta)|^2+8 \sin^2\theta \cos^2\theta |g_3|^2,$$
$$\sigma_{++}(\theta)=\sum_{\lambda_3,\lambda_4}|{\mathcal 
F}_{++,\lambda_3\lambda_4}|^2=\sin^2\theta \left [\displaystyle\frac{9}{2} 
\sin^2\theta |g_2|^2+2(1+\cos^2\theta) |g_3|^2\right ],$$
\begin{eqnarray}
\sigma_{+-}(\theta)=&\sum_{\lambda_3,\lambda_4}|{\mathcal 
F}_{+-,\lambda_3\lambda_4}|^2= 
2|g_1|^2+ 2 Re ~g_1g_2^*(1-3\cos^2 \theta +  \\
&\displaystyle\frac{1}{2} (1-3\cos^2\theta)^2 |g_2|^2+ 2
\sin^2\theta \cos^2\theta |g_3|^2,\nonumber
\end{eqnarray}
$$\sigma_{0+}(\theta)=\sum_{\lambda_3,\lambda_4}|{\mathcal 
F}_{0+,\lambda_3\lambda_4}|^2= 
9\sin^2\theta \cos^2\theta |g_2|^2
+(1-3\cos^2 \theta +4\cos^4\theta)|g_3|^2.$$
With the help of these formulas we can estimate the corresponding integral 
ratios:
$$R_{\lambda_1\lambda_2}=\displaystyle\frac{\int_{-1}^{+1}\sigma_{\lambda_1
\lambda_2}(\theta )d\cos\theta}
{\int_{-1}^{+1}d\cos \theta (d\sigma/d\cos\theta)_0},$$
which characterize the relative role of polarized collisions with respect to 
unpolarized ones:
\begin{equation}
R_{00}=\displaystyle\frac{3}{5}\displaystyle\frac{15+4r}{3+2r},
~R_{++}=\frac{36}{5}\frac{r}{3+2r},~R_{+-}=\frac{12}{5}\frac{15+r}{3+2r},
~R_{0+}=
\displaystyle\frac{9}{5}\displaystyle\frac{r}{3+2r},
\end{equation}
where $r=(3 |g_2|^2+2 |g_3|^2)/|g_1|^2)$. It is interesting that all these 
ratios 
depend on a single contribution of the moduli of the partial amplitudes, the 
ratio 
$r\ge 0$. The ratios $R_{\lambda_1\lambda_2}$ are limited by:
$$1.2\le R_{00}\le 3,~~0\le R_{++}\le 3.6,~~1.2\le R_{+-}\le 12,~~0\le R_{0+}\le 
0.9,$$
where the upper limits correspond to $g_2=g_3=0$, (when only the $g_1$ amplitude 
is 
present), and the lower limits correspond to $g_1=0$ (for any amplitudes $g_2$ 
and 
$g_3$). But the exact values of $R_{\lambda_1\lambda_2}$ depend on the relative 
value 
of the partial amplitudes, through one parameter, $r$.

The general dependence of the differential cross section for $\vec d+\vec 
d$-collisions, can be written in terms of partial cross sections 
$\sigma_{\lambda_1\lambda_2}$ as follows:
\begin{equation}
\displaystyle\frac {d\sigma}{d\Omega}(\vec d+\vec 
d)=(d_+^2+d_-^2)\sigma_{++}(\theta)+d_0^2\sigma_{00}(\theta)+2d_+d_-
\sigma_{+-}(\theta)+2d_0(d_++d_-)\sigma_{0+}(\theta),
\end{equation}
where we used the evident relations between $\sigma_{\lambda_1\lambda_2}$: $
\sigma_{++}(\theta)=\sigma_{--}(\theta)$, 
$\sigma_{0+}(\theta)=\sigma_{0-}(\theta)$, 
$\sigma_{+-}(\theta)=\sigma_{-+}(\theta)$, due to the P-invariance of 
the 
strong interaction, and the standard notation: $d_+$, $d_0$ and  $d_-$ for 
different 
deuteron fractions in polarized plasma.

Using Eq. (20) one can find some interesting limiting cases. Setting for 
example, 
$d_+=d_-$ (deuterons with tensor polarization only: $P_{zz}=1-3d_0$, $P_z=0$), 
one can 
obtain the following dependence of the differential cross section on $P_{zz}$:
\begin{equation}
\displaystyle\frac{d\sigma}{d\Omega}(\vec d+\vec 
d)=a_0(\theta)+2P_{zz}a_1(\theta)+\displaystyle\frac{1}{2}P_{zz}^2a_2(\theta),
\end{equation}
where the coefficients $a_i(\theta),i=0-2$, are linear combinations of the  
helicity 
cross sections $\sigma_{\lambda_1\lambda_2}$:
$$
9a_0(\theta)=2\left[ \sigma_{++}(\theta)+\sigma_{+-}(\theta)\right 
]+\sigma_{00}(\theta)+4\sigma_{+0}(\theta),$$
\begin{equation}
9a_1(\theta)=\sigma_{++}(\theta)+\sigma_{+-}(\theta)-\sigma_{00}(\theta)-\sigma_
{+0}
(\theta),
\end{equation}
$$
9a_2(\theta)=\sigma_{++}(\theta)+\sigma_{+-}(\theta)+2\sigma_{00}(\theta)-
4\sigma_{+0}(\theta).$$
So, measuring the $P_{zz}$-dependence of the cross section for $\vec d+\vec d$ 
collisions, one can determine all 3 coefficients $a_i(\theta)$ (at each angle 
$\theta$). This allows to determine the individual helicity partial cross 
sections  
$\sigma_{\lambda_1\lambda_2}(\theta)$:
$$\sigma_{00}(\theta)=a_0(\theta)-4a_1(\theta)+2a_2(\theta),$$
\begin{equation}
\sigma_{0+}(\theta)=a_0(\theta)-a_1(\theta)-a_2(\theta),
\end{equation}
$$\sigma_{++}(\theta)+\sigma_{+-}(\theta)=2a_0(\theta)+4a_1(\theta)+a_2(\theta).
$$
In order to disentangle the  $\sigma_{++}(\theta)$ and $\sigma_{+-}(\theta)$ 
contributions, an additional polarization observable has to be measured, from 
the 
collisions of vector polarized deuterons ($d_{\pm}=\displaystyle\frac 
{1}{3}\pm\displaystyle\frac {1}{2}P_z,~d_0=\displaystyle\frac{1}{3}$):
\begin{equation}
\displaystyle\frac{d\sigma}{d\Omega}(\vec d+\vec 
d)=a_0(\theta)+\displaystyle\frac{P_z^2}{2}(\sigma_{++}(\theta)-\sigma_{+-}
(\theta)).
\end{equation}

The linear $P_z$ 
contribution is forbidden by the P-invariance of the strong interaction. Only
the measurement of the $P_z^2$ contribution allows to separate the cross 
sections 
$\sigma_{++}(\theta)$ and $\sigma_{+-}(\theta)$.

This analysis is equivalent to the discussion  of the complete experiment (in 
terms of 
helicity cross sections $\sigma_{\lambda_1\lambda_2}(\theta)$).

Finally let us derive the polarization properties of the neutrons in the process 
 $\vec d+\vec d\rightarrow n+^3\!He$. Using eqs. (17) for the helicity 
amplitudes, one 
can find for the $\theta$ dependence of the neutron polarization (for the 
different 
spin configurations of the colliding deuterons):
$$ (n_+-n_-)\sigma_{++}(\theta)=2\sin^2\theta d_+^2\left [ 3 Re 
g_2g_3^*+2\cos^2\theta 
|g_3|^2\right ],$$
\begin{equation}
(n_+-n_-)\sigma_{0+}(\theta)=2d_0d_+ \cos^2\theta\left [ 
-(1-2\cos^2\theta)|g_3|^2+3\sin^2\theta Re~ g_2g_3^*\right ],
\end{equation}
$$(n_+-n_-)\sigma_{00}(\theta)=(n_+-n_-)\sigma_{+-}(\theta)=0,$$
where $n_+$ and $n_-$ are the fractions of polarized neutrons with spin 
parallel and antiparallel relative to the $\vec B$ direction.

The production of unpolarized neutrons for 00-collisions of deuterons results 
from 
P-invariance, and for $-+$ collisions  results from  the identity of colliding 
deuterons and from the P-invariance.
\subsection{\boldmath{Complete experiment for 
$\lowercase{d+d\rightarrow n}+^3\!H\lowercase{e}$}}

Due to three complex partial amplitudes for the S-wave $dd-$interaction for  the 
process
$d+d\rightarrow n+^3\!He$,
 the measurement of a large number of observables is necessary, in order to 
perform  the 
complete 
experiment. This study  will be based on the formalism of the 
polarized structure functions, previously used in \cite{Rek98} for the process 
$d+^3\!H\rightarrow n+^4\!He$.

Let us consider the collisions of polarized deuterons 
$\vec d+\vec d\rightarrow n+^3\!He$. The differential cross section can be 
parametrized in the following general form:
\begin{equation}
\begin{array}{ll}
\displaystyle\frac{d\sigma}{d\Omega}=&\left 
(\displaystyle\frac{d\sigma}{d\Omega}\right 
)_0\left [ 1+{\mathcal A}_1(\vec k\cdot\vec Q_1+\vec k\cdot \vec Q_2)+{\mathcal 
A}_2 \vec S_1\cdot \vec S_2+{\mathcal A}_3\vec k\cdot\vec S_1~\vec k\cdot\vec 
S_2\right 
.\\
& +{\mathcal A}_4\vec k\cdot\vec Q_1~\vec k\cdot \vec Q_2+{\mathcal A}_5\vec 
Q_1\cdot \vec Q_2+{\mathcal A}_6 Q_{1ab}Q_{2ab}\\
&\left .
+{\mathcal A}_7(\vec k\cdot\vec S_1\times\vec Q_2+\vec k\cdot\vec S_2\times\vec 
Q_1)\right 
], ~Q_{1a}=Q_{1ab}k_b,~Q_{2a}=Q_{2ab}k_b,
\end{array}
\end{equation}
where $\vec S_1$ and $\vec S_2$ ($Q_{1ab}$ and $Q_{2ab}$) are the vector 
(tensor) 
polarizations of the colliding deuterons. The real coefficient ${\mathcal A}_1$ 
describes 
the tensor analyzing power in $\vec d+ d\rightarrow n+^3\!He$, ${\mathcal 
A}_2-{\mathcal A}_7$ are the spin 
correlation 
coefficients in $\vec d+\vec d\rightarrow n+^3\!He$. The coefficients ${\mathcal 
A}_1-{\mathcal 
A}_6$ are T-even polarization observables and ${\mathcal A}_7$ is the T-odd one 
(due to 
the specific correlation of the vector polarization of one deuteron and the 
tensor 
polarization of the other deuteron). Note that these coefficients ${\mathcal 
A}_i$ 
can not 
fix the relative phases of the singlet amplitudes $g_1$ and $g_2$ (from 
one side) 
and the triplet amplitude $g_3$ (from the other side). The complete experiment 
has to 
be more complex than the determination of the polarization observables 
${\mathcal 
A}_i$. 
The polarization transfer coefficients from the initial deuteron to the produced 
fermion ($n$ or $^3H$) have to be measured, too.

After summing over the polarizations of the produced particles in $\vec d+\vec 
d\rightarrow n+^3\!He$, the following expressions can be found, for the 
coefficients 
${\mathcal A}_i,~i=1-7$, in terms of the partial amplitudes $g_k,~k=1-3$:
$$
-\displaystyle\frac{9}{2}{\mathcal A}_1\left (\displaystyle\frac 
{d\sigma}{d\Omega}\right)_0=3|g_2|^2+|g_3|^2+6 Re~ g_1g_2^*,$$
$$
 {\mathcal A}_2\left (\displaystyle\frac{d\sigma}{d\Omega}\right)_0= 
-|g_1|^2+2|g_2|^2+|g_3|^2- Re~ g_1g_2^*,$$
 $$
{\mathcal A}_3\left(\displaystyle\frac{d\sigma}{d\Omega}\right)_0=
-3|g_2|^2-|g_3|^2+3Re~ g_1g_2^*,$$
\begin{equation}
 \displaystyle\frac{9}{4}{\mathcal A}_4
\left 
(\displaystyle\frac{d\sigma}{d\Omega}\right)_0= 9|g_2|^2-4|g_3|^2
\end{equation}
$$
\displaystyle\frac{9}{2}{\mathcal A}_5\left (\displaystyle\frac 
{d\sigma}{d\Omega}\right)_0= -6|g_2|^2+ 6Re~ g_1g_2^*+2|g_3|^2,
$$
$$
 \displaystyle\frac{9}{2}{\mathcal A}_6\left (\displaystyle\frac 
{d\sigma}{d\Omega}\right)_0= |g_1|^2+|g_2|^2-2Re~ g_1g_2^*,$$
$$
{\mathcal A}_7 \left 
(\displaystyle\frac{d\sigma}{d\Omega}\right)_0=-2~Im ~g_1g_2^*,
$$
where $({d\sigma}/{d\Omega})_0$ is the differential cross section with 
unpolarized 
particles:
$$\left (\displaystyle\frac {d\sigma}{d\Omega}\right )_0=\frac{2}{9}\left [ 3 
|g_1|^2+6|g_2|^2+4|g_3|^2\right ]=\displaystyle\frac{2}{9}|g_1|^2(3+2r).$$
Using these expressions, the following relations can be found between the 
coefficients 
${\mathcal A}_i$:
\begin{itemize}
\item [(a)] linear: between T-even polarization observables,
$${\mathcal A}_2+{\mathcal A}_3+\displaystyle\frac{9}{2}{\mathcal A}_6
={\mathcal A}_1+{\mathcal A}_4-\displaystyle\frac{1}{3}
{\mathcal A}_3+\displaystyle\frac{7}{4}{\mathcal A}_5=0$$
\item[(b)] quadratic, relating the T-odd asymmetry ${\mathcal A}_7$ with the 
T-even 
coefficients ${\mathcal A}_i,~i=1-6$;
\end{itemize}
$$ \displaystyle\frac{9}{4}(1+{\mathcal A}_1^2-{\mathcal A}_7^2)
={\mathcal A}_2^2+({\mathcal A}_2+{\mathcal A}_3)^2
+6({\mathcal A}_1{\mathcal A}_2+{\mathcal A}_1{\mathcal A}_3+{\mathcal 
A}_2{\mathcal A}_3)$$

Therefore, the measurements of $({d\sigma}/{d\Omega})_0$ and 3 coefficients 
${\mathcal 
A}_i,~i=1-3$, allow to find the moduli of all S-wave partial amplitudes 
$g_k,~k=1-3$, and the relative phase of the singlet amplitudes $g_1$ and $g_2$:
$$18|g_1|^2=(9-12{\mathcal A}_2-4{\mathcal A}_3)\left (\displaystyle\frac 
{d\sigma}{d\Omega}\right )_0,$$
$$-18|g_2|^2=(9+18{\mathcal A}_1+ 6{\mathcal A}_2+10{\mathcal A}_3)\left 
(\displaystyle\frac 
{d\sigma}{d\Omega}\right )_0,$$
$$2|g_3|^2=(3+3{\mathcal A}_1+ 2{\mathcal A}_2+2{\mathcal A}_3)\left 
(\displaystyle\frac 
{d\sigma}{d\Omega}\right )_0,$$
$$18 Re~g_1g_2^*=(-9{\mathcal A}_1+2{\mathcal A}_3)\left (\displaystyle\frac 
{d\sigma}{d\Omega}\right )_0.$$
So these measurements can be considered as the first step of the complete 
experiment 
for the process 
$ d+d\rightarrow n+^3\!He$ in the  near threshold conditions.

Using these expressions, one can find the following expression for the ratio 
$r$:
$$ r=3\displaystyle\frac{1+a}{1-2a},~a=
\displaystyle\frac{2}{9}(3{\mathcal A}_2+{\mathcal 
A}_3).$$

Let us discuss, for completeness, the simplest polarization phenomena for the 
inverse 
reaction, $ n+^3\!He\rightarrow d+d$, when the deuterons are produced in 
S-state. In 
this case this process can be described by the same set of partial amplitudes. 
The 
cross section for the collisions of polarized particles, $ \vec n+\vec 
{^3\!He}\rightarrow d+d$, can be parametrized in the following way:
\begin{equation}
\displaystyle\frac{d\sigma}{d\Omega}(\vec n+\vec {^3\!He})=\left 
(\displaystyle\frac{d\sigma}{d\Omega}\right)_0\left [1+B_1\vec P_1\cdot\vec 
P_2+B_2\vec 
k\cdot\vec P_1~\vec k\cdot\vec P_2\right ],
\end{equation}
where $\vec P_1$ and $\vec P_2$ are the polarizations for $n$ and $^3\!He$ and 
$B_i$, i=1,2, are the spin correlation coefficients for the inverse reaction, 
which can be expressed as functions of the partial amplitudes:
\begin{equation}
B_1=-3\displaystyle\frac{|g_1|^2+2|g_2|^2}{3|g_1|^2+6|g_2|^2+4|g_3|^2},~~B_2=
\displaystyle\frac{4|g_3|^2}{3|g_1|^2+6|g_2|^2+4|g_3|^2}.
\end{equation}
These coefficients are not independent, $-B_1+B_2=1$, as a result of the 
peculiar spin 
structure of the matrix element for the $n+^3\!He-$collision. Comparing Eq. (27) 
and Eq. (29), one can 
find the 
following relation:
$$\displaystyle\frac{9}{4}B_2=3(1+{\mathcal A}_1)+2({\mathcal A}_2+{\mathcal 
A}_3).$$

In order to find the relative phases of the singlet and triplet amplitudes in 
$ d+ d\rightarrow n+^3\!He$, it is necessary to measure the polarization 
transfer 
coefficients (from the initial deuteron to the final neutron). The most general 
parametrization of the neutron polarization can be written in the following 
form:
$$\vec P=p_1\vec S+p_2\vec k(\vec k\cdot\vec S) +p_3\vec k\times\vec Q,$$
where $p_i$, i=1-3, are the real structure functions, characterizing the 
corresponding 
coefficients of polarization transfer:
$$p_1=p_2=0,~~p_3=-\displaystyle\frac{4Im(g_1-g_2)g_3^*}{3|g_1|^2+6|g_2|^2+4|g_3
|^2}.$$
Both T-even coefficients $p_1$ and $p_2$ are identically zero. This results from 
the specific spin 
structure of the 
threshold 
amplitude, due to the identity of the colliding 
deuterons 
and the S-wave interaction. The T-odd coefficient $p_3$ is sensitive to the 
relative 
phases of singlet and triplet amplitudes.
 
\section{\boldmath{$^3\!H\lowercase{e}$ and $^3\!H$ collisions}}

\subsection{Three-particle production}
We discuss here the spin structure and the polarization observables for the 
following  processes (at low energy for colliding particles):
$$ ^3\!H+^3\!H \rightarrow 2n+^4\!He,~~^3\!H+^3\!He \rightarrow n+p+^4\!He$$
$$ ^3\!He+^3\!He \rightarrow 2p+^4\!He,~~^3\!H+^3\!He \rightarrow d+^4\!He$$
Due to the presence of identical fermions in initial and final states, the spin 
structure of the symmetric processes, $ ^3\!He+^3\!He \rightarrow 
2p+^4\!He$ 
and 
$ ^3\!H+^3\!H \rightarrow 2n+^4\!He$, for the S-state interaction of colliding 
nuclei is built on a single transition: $S_i=S_f=0$, with the following 
parametrization:
$$
g_0(\chi_2^{\dagger}\sigma_2\widetilde{\chi_1}^{\dagger})(\widetilde{\phi_1}
\sigma_2\phi_2),
$$ 
where $\chi_1$ and $\chi_2$ ($\phi_1$ and $\phi_2$) are the 2-component 
spinors of 
the produced nucleons (colliding nuclei), $g_0$ is the singlet-singlet 
amplitude, which 
dependence on the energies of the produced particles has a dynamical character 
and can 
be, in principle, complicated. However a model independent expression  of the 
cross 
section as a function of the polarizations $\vec P_1$ and $\vec P_2$ of the 
colliding 
nuclei can be given as: $ 
\displaystyle\frac{d\sigma}{d\omega}(\vec P_1,\vec P_2)=\left 
(\displaystyle\frac{d\sigma}{d\omega}\right)_0(1-\vec P_1\cdot\vec P_2).$
This dependence is correct for any 3-particle phase-space element $d\omega$, and 
then, 
also for the total cross section. Therefore the polarization of colliding nuclei 
decreases the reaction rate, in comparison with unpolarized collisions. In a 
fusion reactor, this property will favour the plasma production through  
the main 
$d+^3\!H$-reaction, which has a larger Q-value and uses $d-$fuel, preventing the 
waste of the very expensive $^3\!H$ in 'non effective' $^3\!H+^3\!H$ collision. 

Due to the non-identity of the colliding and produced fermions, the process 
$^3\!H+^3\!He \rightarrow n+p+^4\!He$ is characterized by two independent 
transitions, 
a singlet one, (with amplitude equal to half of the $^3\!H+^3\!H$ amplitude, due 
to the isotopic invariance of the nuclear interaction) and a 
triplet amplitude. then the matrix element can be written as:
\begin{equation}
{\mathcal M}(^3\!H+^3\!He)=
\displaystyle\frac{1}{2}g_0(\chi_2^{\dagger}\sigma_2\widetilde{\chi_1}^{\dagger}
)
(\widetilde{\phi_1}\sigma_2\phi_2)+\displaystyle\frac{1}{2}g_1(\chi_2^{\dagger}
\sigma_a\sigma_2\widetilde{\chi_1}^{\dagger})(\widetilde{\phi_1}\sigma_2\sigma_a
\phi_2).
\end{equation}

Therefore the dependence of the cross section for 
$\vec{^3\!H}+\vec{^3\!He}$-collisions 
on the polarizations of the colliding particles is:
$$\displaystyle\frac{d\sigma}{d\omega}(\vec P_1,\vec P_2)=\left 
(\displaystyle\frac{d\sigma}{d\Omega}\right)_0(1+{\mathcal A}\vec P_1\cdot\vec 
P_2),~~{\mathcal 
A}=\displaystyle\frac{|g_0|^2+|g_1|^2}{|g_0|^2+3|g_1|^2},~~-1\le {\mathcal A}\le 
\displaystyle\frac{1}{3}.$$
The coefficient ${\mathcal A}$ is determined by the ratio $R$ of the cross 
sections 
for 
$^3\!H+^3\!H$ and $^3\!H+^3\!He$ collisions (with unpolarized particles):
$$R=\displaystyle\frac{\sigma(^3\!H+^3\!H)}{\sigma(^3\!H+^3\!He)}=\displaystyle
\frac{4|g_0|^2}{(|g_0|^2+3|g_1|^2)},~6{\mathcal A}=2+R.$$
This result, of course, is valid only for the S-state interaction 
in the considered reactions.

The relative phase of the partial amplitudes $g_0$ and $g_1$ determines the 
polarization transfer from the initial ${^3\!He}$ to the final $n$:
$$\vec P_n=p\vec P_i,~p=2\displaystyle\frac{|g_1|^2+Re~ 
g_0g_1^*}{|g_0|^2+3|g_1|^2}.$$

Note that in the process $^3\!H+^3\!He$ we neglected the production of final 
particles 
in the D-state, which is allowed in the general case by the conservation of 
the total angular momentum and P-parity: $S_i=1\rightarrow S_f=1,~\ell_{1,2}=2$, 
where $\ell_1$ is the orbital angular momentum of the $np-$system and $\ell_2$ 
is the orbital angular momentum 
of the 
$ ^4\!He$, relative to the center of mass of the $np$-system. The spin structure 
for these transitions can be written as:
$$\chi_2^{\dagger}(k_a\vec\sigma\cdot\vec 
k-\displaystyle\frac{1}{3}\sigma_a)\sigma_2\widetilde{\chi_1}^{\dagger}~
(\widetilde{\phi_1}
\sigma_2\sigma_a\phi_2),$$
where $\vec k$ is the unit vector along the 3-momentum of the $^4\!He$ (if 
$\ell_{2}=2,~ \ell_{1}=0$) or along the proton 3-momentum (if 
$\ell_{1}=2,~\ell_{2}=0$).
The matrix element for the production of two P-waves, $\ell_{1}=\ell_{2}=1$ is 
comparable, in principle, to the D-wave production.
\subsection{The deuteron production}
Let us finally discuss the process $^3\!H+^3\!He \rightarrow d+^4\!He$, 
(Q-value=14.3 MeV), which is also occurring in a $d+d$ or $d+^3\!He$-fusion 
reactor. The 
spin structure of the threshold amplitude can be established using the 
generalized 
Pauli principle (which is valid at the level of the isotopic invariance). Due to 
the isospin value $I=0$ for the entrance channel, the sum 
$S_i+\ell_i$ must be $odd$, and for S-interaction ($\ell_i=0$), we have $S_i=1$, 
with 
the following two transitions allowed:
$S_i=1~\rightarrow {\mathcal 
J}^{\pi}=1^+~\rightarrow ~{\ell}_f=0$ and ${\ell}_f=2$, where ${\ell}_f$ is the 
orbital 
angular momentum of the produced deuteron. Therefore, the matrix element can be 
written 
as:
\begin{equation}
{\mathcal M}=\widetilde{\phi_1}\sigma_2\left[h_0\vec\sigma\cdot 
\vec 
D^*+h_2 
(3\vec\sigma\cdot \vec k~\vec k\cdot \vec D^*-\vec\sigma\cdot\vec D^*)\right ] 
\phi_2,
\end{equation}
where $\vec D$ is the deuteron polarization 3-vector, $h_0$ and $h_2$ are the 
partial 
amplitudes, corresponding to ${\ell}_f=0$ and 2. These amplitudes are complex 
functions 
of the excitation energy. The complete experiment, for this process, must 
contain the 
measurement of at least three different observables. 

The dependence of the 
differential 
cross section on the polarizations of the colliding nuclei, have the same 
structure as 
eq. (28), with:
$$B_1\left 
(\displaystyle\frac{d\sigma}{d\Omega}\right)_0=\displaystyle\frac{1}{2}|h_0|^2+2
Re~h_0h
_2^*+2|h_2|^2,~~B_2\left 
(\displaystyle\frac{d\sigma}{d\Omega}\right)_0=-3(|h_2|^2+2Re~h_0h_2^*),$$
$$\left 
(\displaystyle\frac{d\sigma}{d\Omega}\right)_0=\displaystyle\frac{3}{2}(|h_0|^2+
2|h_2|^
2),~~3 B_1+ B_2=1.$$ This results from the absence of a singlet state of the 
colliding nuclei, in the discussed process.

After $\vec k$- integration, one can find:
$$\sigma(\vec P_1,\vec P_2)=\sigma_0(1+B\vec P_1 \cdot \vec 
P_2),~B=B_1+\displaystyle\frac{1}{3}B_2=\displaystyle\frac{1}{3},$$
independently on the relative value of the partial amplitudes. Therefore the 
dependence 
of the total cross section from the polarizations can be predicted exactly, due 
to the 
specific spin structure of the threshold matrix element.

The information that can be obtained from the tensor analyzing power in the 
inverse 
reaction,  $\vec d+^4\!He \rightarrow ^3\!H+^3\!He$, does not give any 
new constraint on the partial amplitudes. The cross section is written, in terms 
of 
the 
tensor analyzing power ${\mathcal A}_d$:
$$\displaystyle\frac{d\sigma}{d\Omega}(\vec d+^4\!He )=\left 
(\displaystyle\frac{d\sigma}{d\Omega}\right)_0\left [1+{\mathcal A}_d 
Q_{ab}k_ak_b\right 
]$$ and 
$${\mathcal A}_d= -\displaystyle\frac{|h_2|^2+2Re~h_0h^*_2}{|h_0|^2+2|h_2|^2},$$
i. e. ${\mathcal A}_d=\displaystyle\frac{1}{3}B_2$. This means that the 
measurement of the 
tensor 
analyzing 
power ${\mathcal A}_d$ allows to define both correlation coefficients, for the 
direct 
reaction $\vec{^3\!H}+  \vec{^3\!He}\rightarrow d+^4\!He$. 

The complete 
experiment  has to include the measurement of polarization transfer 
coefficients in 
$^3\!H+\vec{^3\!He}\rightarrow {\vec d}+^4\!He$, where the vector deuteron 
polarization 
$\vec 
S$ can be written in the following general form: 
$$\vec S=s_1\vec P+s_2\vec k~\vec k\cdot\vec P,$$
with the following expressions for the real coefficients $s_i,~i=1,2:$
$$s_1=\displaystyle\frac{|h_0|^2-Re~h_0h^*_2}{|h_0|^2+2|h_2|^2},~~s_2=
\displaystyle\frac{-Re~h_0h^*_2+|h_2|^2}{|h_0|^2+2|h_2|^2}.$$
So, finally, the complete experiment for $^3\!H+{^3\!He}\rightarrow d+^4\!He$ 
has to include 
the 
measurements of $({d\sigma}/{d\Omega})_0$, $B_1$ and $s_1$:
$$9|h_0|^2=2(3B_1+4s_1-2)\left (\displaystyle\frac{d\sigma}{d\Omega}\right)_0,$$
$$-9|h_2|^2=(3B_1+4s_1-5)\left (\displaystyle\frac{d\sigma}{d\Omega}\right)_0,$$
$$-9Re~h_0h^*_2=2(-3B_1-s_1+2)\left 
(\displaystyle\frac{d\sigma}{d\Omega}\right)_0.$$
The results of all other polarization experiments can be expressed in terms of 
these 
quantities. As an example, let us consider the tensor deuteron polarization from 
$^3\!H+\vec {^3\!He}\rightarrow \vec d+^4\!He$:
$$Q_{ab}=Q\left [ k_a(\vec k\times\vec P)_b+k_b(\vec k\times
\vec P)_a\right ],~Q=\displaystyle\frac{Im~h_0h^*_2}{|h_0|^2+2|h_2|^2}.$$
This T-odd polarization observable can be connected with the T-even coefficients 
$s_1$ 
and $B_1$ through the quadratic relation:
$$\displaystyle\frac{9}{2}Q^2=-B_1(B_1-1)-2(1-s_1-B_1)^2.$$
In order to analyze the angular dependence of the reaction products for 
$\vec{^3\!H}+\vec {^3\!He}\rightarrow d+^4\!He$ in a magnetic field $\vec B$, 
one can 
extend the formalism of helicity amplitudes, as for the previously studied 
reactions.

\section{Conclusions}
We have studied nuclear reactions, at low energies, involved in 
fusion reactors, 
with special emphasis  on their strong dependence on the polarization states of 
the colliding nuclei. The results obtained here on the angular dependence and 
the reaction rate dependence on the polarizations, can be used as a guideline in 
the conception of magnetic fusion reactors. The polarization of the produced 
particles is also important, as it can help the fusion process in a working 
reactor. 
For example, in a a reactor based on $d+^3\!H$-fuel, the intensive flux of 14 
MeV 
neutrons can be used in the $Li-$blanket, not only for its heating, with 
consequent production of electric power, but also to produce extra $^3\!H$- 
fuel, through the processes: 
$n+^6\!Li\rightarrow ^3\!H+^4\!He \mbox{~and ~}n+^7\!Li\rightarrow 
n+^3\!H+^4\!He.$
Due to the definite polarization properties of these reactions, 
one can increase, in principle, the yield of $^3\!H$.

We showed that the polarization and the angular distribution of the neutrons, 
produced in the process $d+^3\!H\rightarrow n+^4\!He$ depends strongly on the 
relative value of the two possible partial amplitudes. The presence of a 
contribution (even relatively small) of the  ${\mathcal J}^\pi =1/2^+$ amplitude 
is 
very important for polarization phenomena.

For the reaction $d+d\rightarrow n+^3\!He$ (with three independent threshold 
partial amplitudes) the situation is more complicated. The $d+d$-reactions 
produce energetic  neutrons and tritium, and should be suppressed in a 
$d+^3\!He$ reactor. 

The detailed information about partial amplitudes of different reactions can be 
obtained, in a model independent way, through the realization of the complete 
experiment. Even at low energy, where the spin structure of all matrix 
elements is highly simplified, the complete experiment includes the scattering 
of a polarized beam on a polarized target.  These experiments, which are absent
up to now,  allow the full reconstruction of the spin structure of the threshold 
amplitudes.

The main results contained in this paper can be summarized as follows:
\begin{itemize}
\item We give a model independent parametrization of the spin structure of the 
threshold matrix elements for the following reactions: $d+d \rightarrow 
n+^3\!He$, 
$d+^3\!H\rightarrow n+^4\!He$, $^3\!H+^3\!H\rightarrow n+n+^4\!He$, 
$^3\!He+^3\!He\rightarrow p+p+^4\!He$, $^3\!H+^3\!He\rightarrow p+n+^4\!He$,
and $^3\!H+^3\!He\rightarrow d+^4\!He$.
\item The helicity amplitudes for the processes $d+d \rightarrow n+^3\!He$ and
$d+^3\!H\rightarrow n+^4\!He$
 are calculated in terms of partial threshold amplitudes, choosing the direction 
of the polarized magnetic field as the quantization axis.
\item The angular distribution of the reaction products for $\vec d+\vec d$ and 
$\vec d+\vec {^3\!He}$-collisions shows a strong dependence on the polarization 
of the colliding particles, 
and 
it can be very important to optimize the blanket and the shielding of a reactor.
\item The polarization properties of neutrons, produced in the processes 
$d+^3\!H$ $\rightarrow $$n+^4\!He$ and $d+d $$\rightarrow$$ n+^3\!He$  are 
derived for 
collisions of polarized particles.
\item The spin structure of the matrix elements and polarization properties are 
derived forsome  A=3 induced processes.
\end{itemize}
\section{Acknowledgements}
We are grateful to E. J. Ludwig and H. J. Karkowski for very fruitful 
discussions.
We thank the Institute for Nuclear Theory at the University of Washington for 
its hospitality, during the completion of this work, and the Department of 
Energy for partial support.

\newpage
\begin{center}
{\large\bf Figure Caption}
\end{center}
\vspace*{2truecm}

{\noindent\bf Figure 1.} Ratio 
$\sigma_{0+}/\sigma_{00}$, as a function of  $x=|g_s|/|g_d|$ and $y=\cos\theta$ 
for the reaction 
$\vec d+\vec {^3\!He}\rightarrow n+^4\!He$,for different values of the phase 
$\delta$: (a) 
$\delta=0$, (b) $\delta=\frac{\pi}{2}$ and (c) $\delta={\pi}$, from Eq. (8).

\vspace*{.5truecm}

{\noindent\bf Figure 2.} Ratio $\sigma_{+-}/\sigma_{00}$ as a function of  
$x$ and $y$ for the reaction 
$\vec d+\vec {^3\!He}\rightarrow n+^4\!He$, for different values of the phase 
$\delta$: (a) 
$\delta=0$, (b) $\delta=\frac{\pi}{2}$ and (c) $\delta={\pi}$, from Eq. (10).
\vspace*{.5truecm}

{\noindent\bf Figure 3.} Neutron polarization  $P_n=\left 
[-2+3y^2+x\cos\delta(1-3y^2) +x^2\right ]/(2+x^2),$  as a function of  
$x$ and $y$ in 
$\vec d+^3\!He$-collisions, for different values of the phase 
$\delta$: (a) $\delta=0$, (b) $\delta=\frac{\pi}{2}$ and (c) $\delta={\pi}$.
\vspace*{.5truecm}

{\noindent\bf Figure 4.} Neutron polarization 
$P_n=\displaystyle\frac{1}{3}
\left [2(2-3y^2)+4x\cos\delta(1-3y^2) +x^2\right ]/(2+x^2)$, as a function of  
$x$ and $y$ in 
unpolarized $d+^3\!He$-collisions,  for different values of the phase 
$\delta$: (a) $\delta=0$, (b) $\delta=\frac{\pi}{2}$ and (c) $\delta={\pi}$.
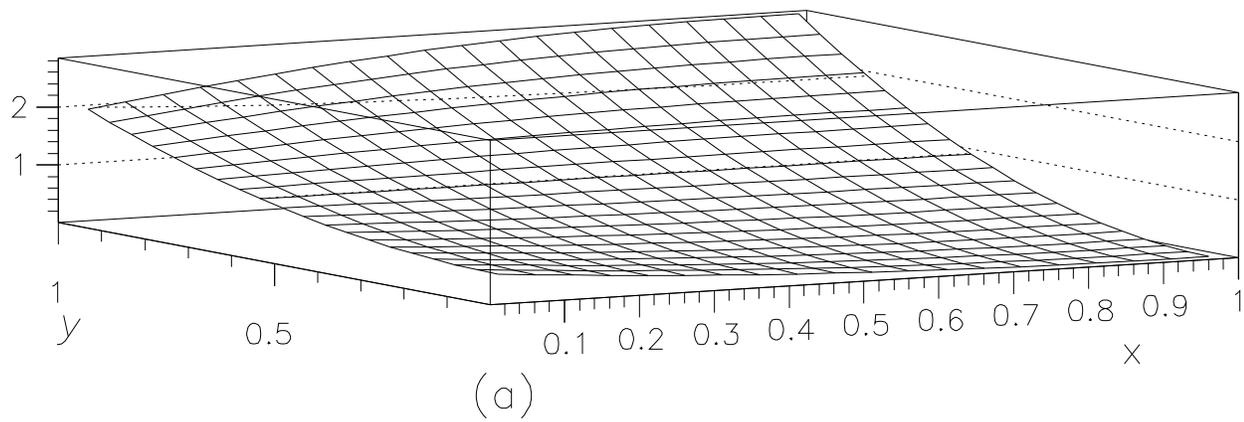
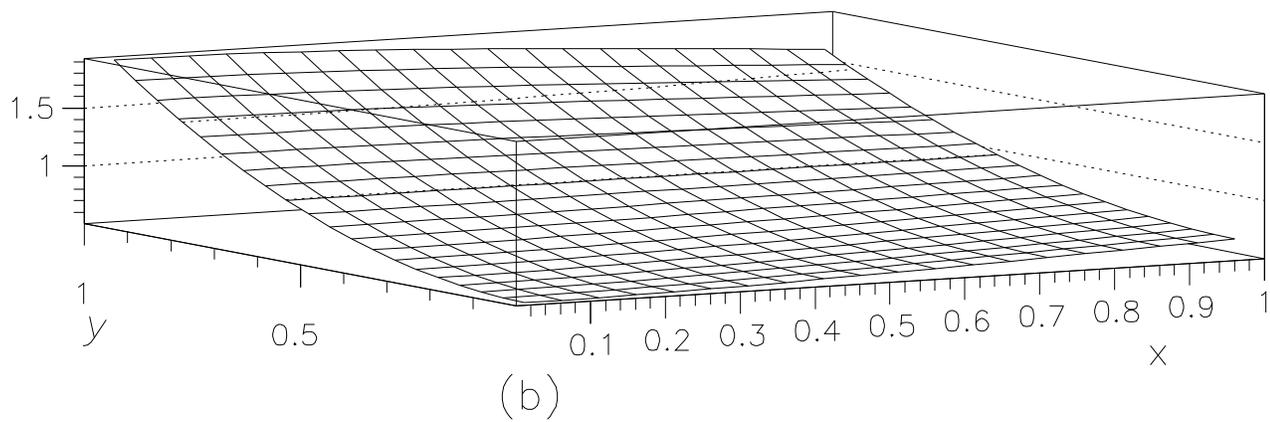
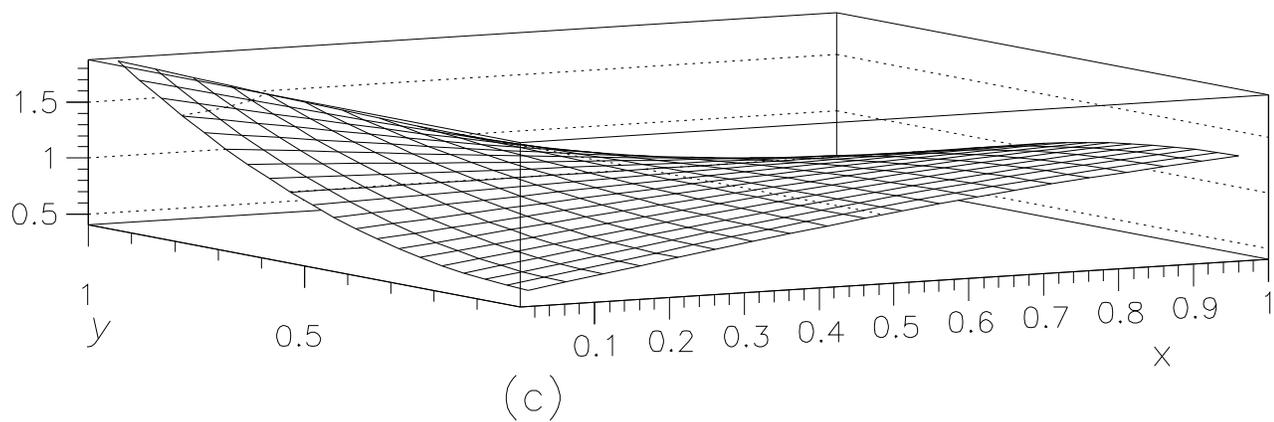
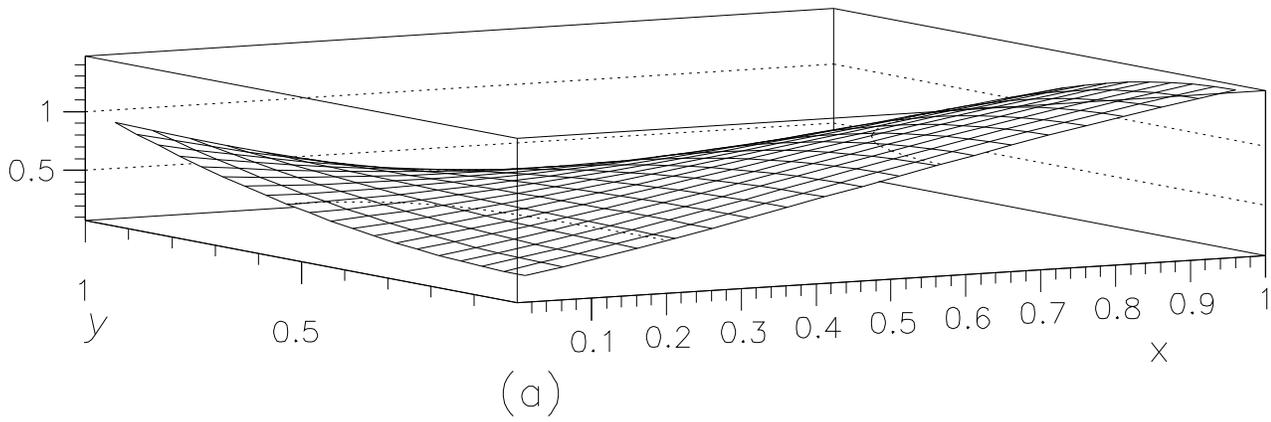
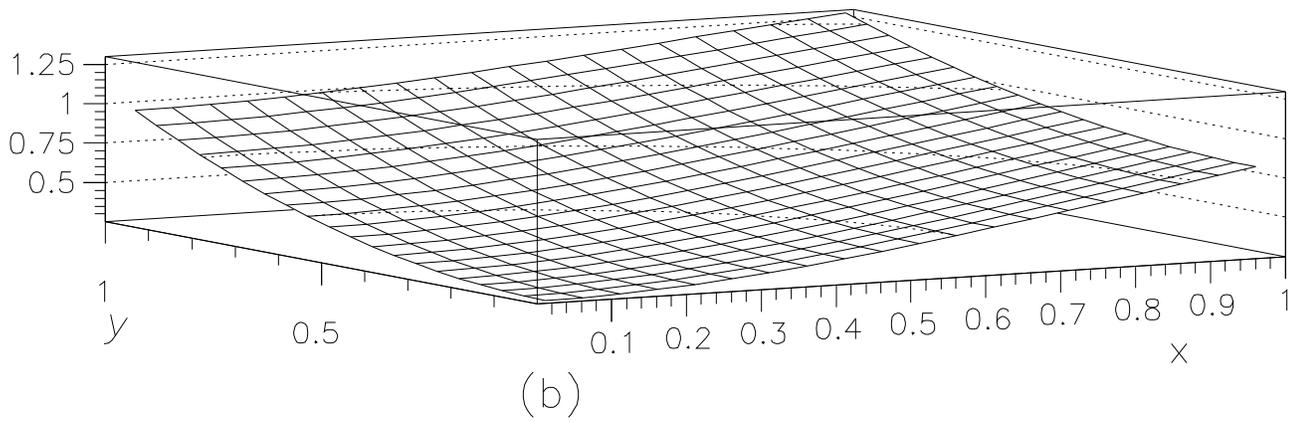
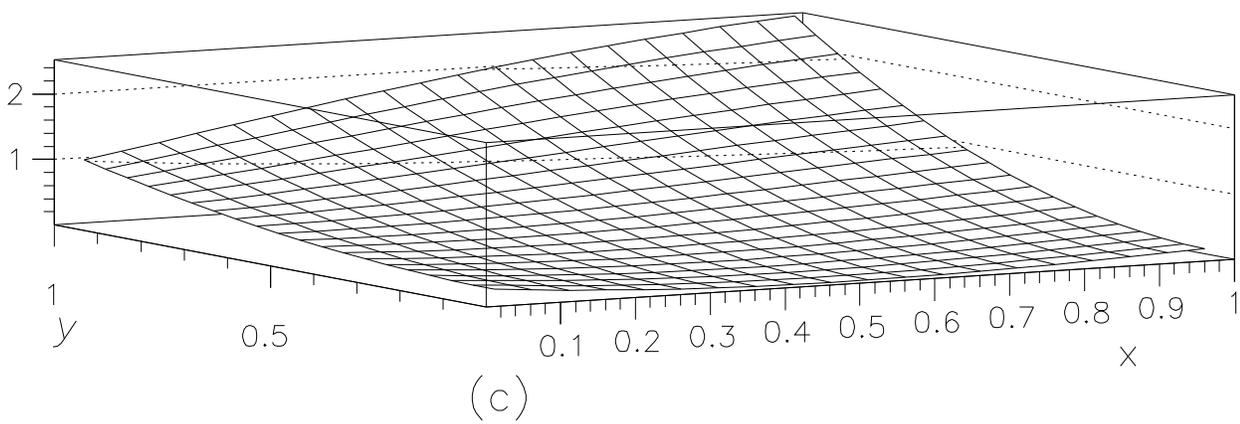
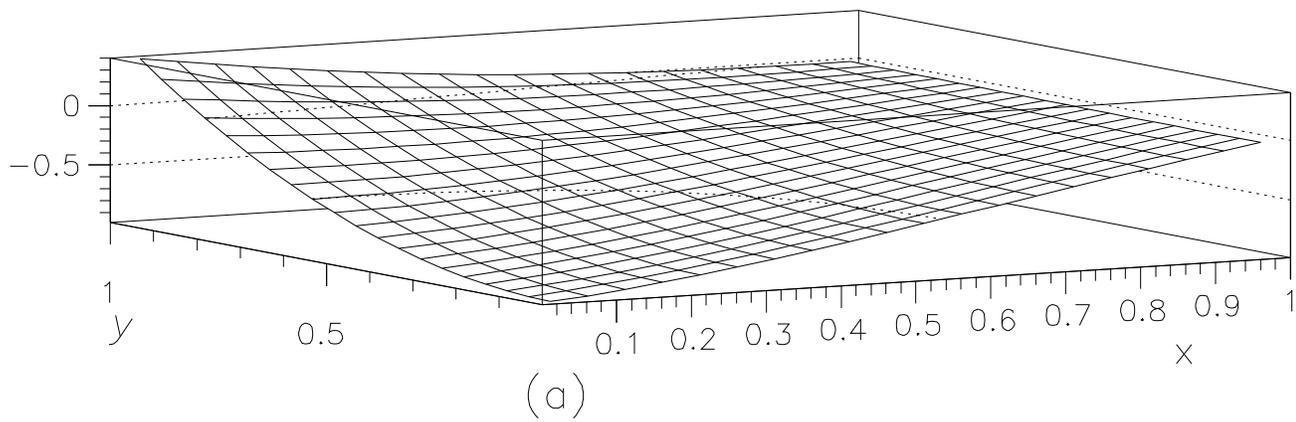
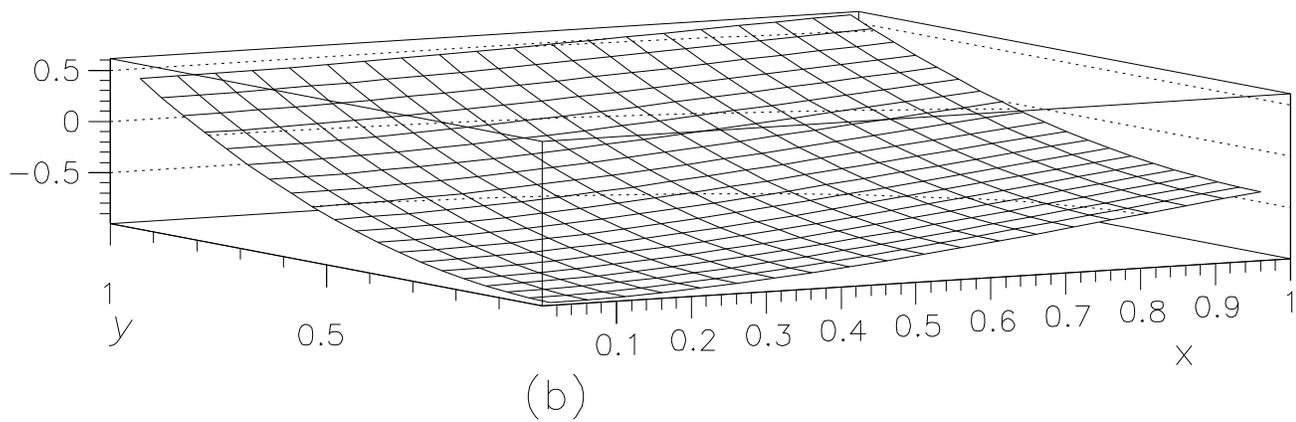
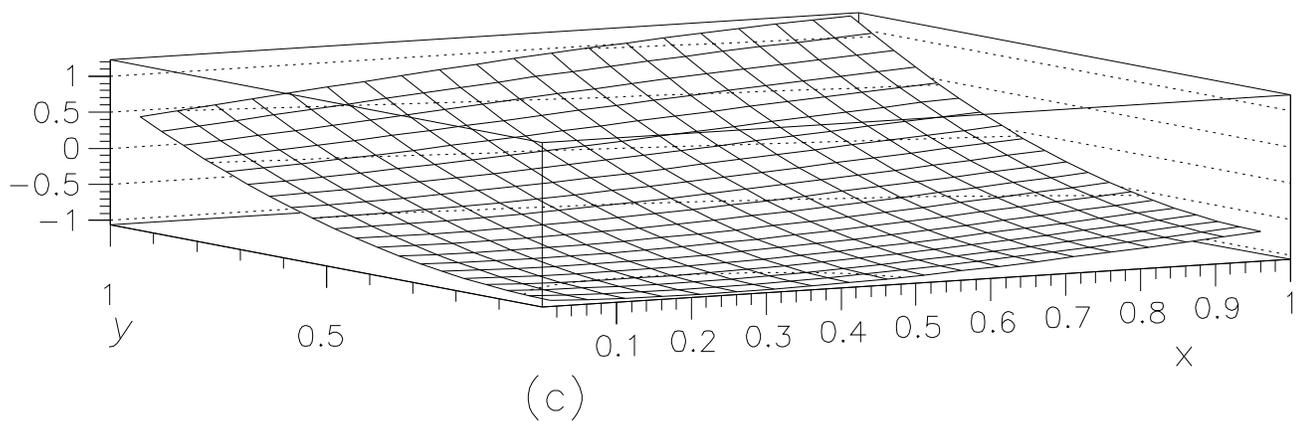
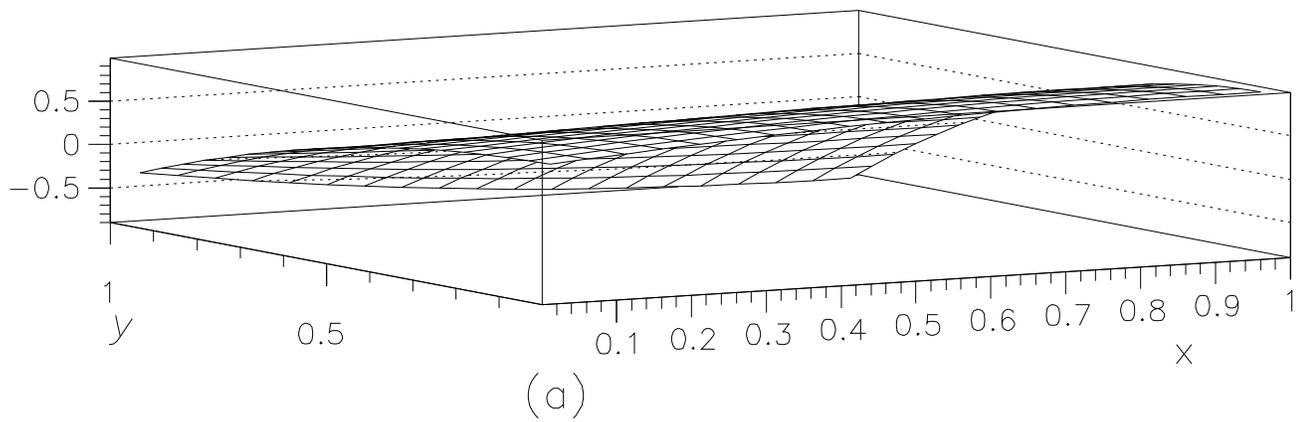
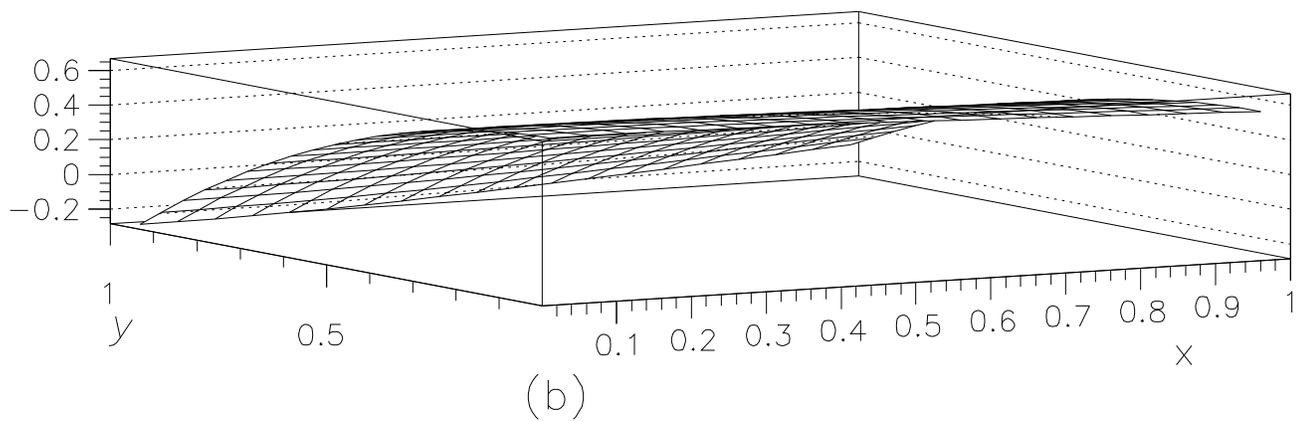
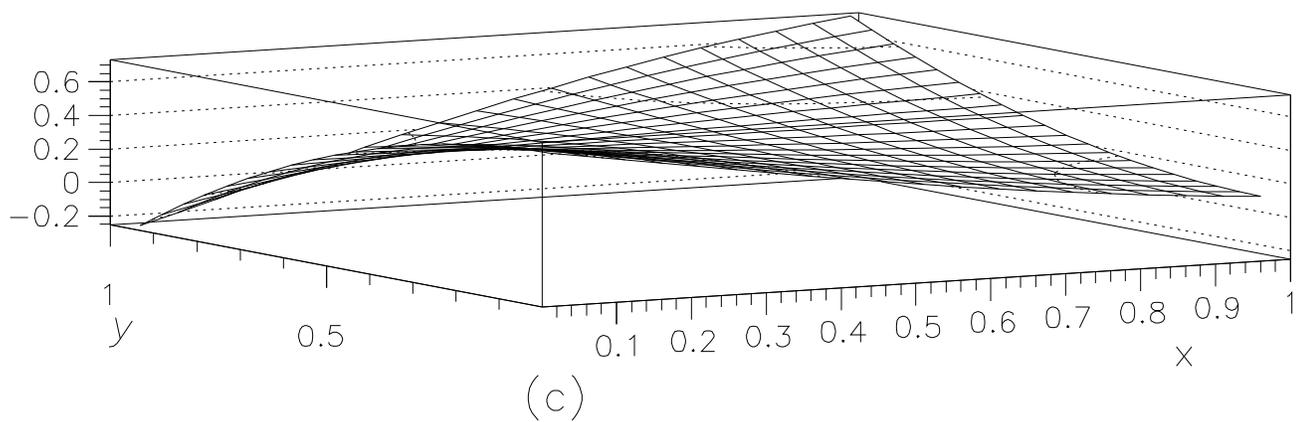
\end{document}